\documentclass[conference]{IEEEtran}
\IEEEoverridecommandlockouts
\usepackage{cite}
\usepackage{amsmath,amssymb,amsfonts}
\usepackage{algorithmic}
\usepackage{graphicx}
\usepackage{textcomp}
\usepackage{xcolor}
\def\BibTeX{{\rm B\kern-.05em{\sc i\kern-.025em b}\kern-.08em
    T\kern-.1667em\lower.7ex\hbox{E}\kern-.125emX}}
\begin{document}

\title{Deterministic Runtime-Adaptive CAN Routing Diversity for Replay Attack Mitigation}

\author{\IEEEauthorblockN{1\textsuperscript{st} Ashwin Kulkarni}
\IEEEauthorblockA{\textit{TORC Robotics, Inc.} \\
\textit {Ann Arbor, MI, USA}\\
}
\and
\IEEEauthorblockN{2\textsuperscript{nd} Aditi Ramadwar}
\IEEEauthorblockA{\textit{Autonomous Driving Department} \\
\textit{Mercedes-Benz Research and Development North America}\\
\textit{San Jose, CA, USA}\\
}
}

\maketitle

\begin{abstract}
Controller Area Network (CAN) remains the dominant in-vehicle communication protocol for real-time control, but its broadcast architecture and lack of native authentication or freshness protection make replay and spoofing attacks feasible once bus access is obtained. This paper proposes a deterministic runtime-adaptive routing overlay that introduces logical route diversity without modifying the CAN physical layer or arbitration mechanism. Participating Electronic Control Units (ECUs) maintain multiple predefined routing patterns and independently select the active route using a shared deterministic function of synchronized, quantized runtime vehicle inputs and a time reference. The resulting coordinated route switching creates time-varying message propagation paths that reduce the validity of previously recorded traffic. A simulation study evaluates routing-index entropy, routing-index repetition behavior, replay acceptance, and false rejection under synchronization jitter. Results show that, with 500 routing patterns, routing-index entropy approaches the theoretical maximum while, under the evaluated simulation conditions, replay acceptance for the proposed overlay is approximately 0.2\% with strict synchronization and remains below 3\% under relaxed synchronization tolerances. The static-routing baseline is a comparative simulation construct only and does not describe the cybersecurity architecture of any production vehicle or account for independent protections that may be present in fielded systems. The proposed approach is not intended to replace cryptographic authentication mechanisms such as AUTOSAR SecOC, but rather to provide an additional structural-diversity layer within a defense-in-depth automotive cybersecurity architecture.
\end{abstract}


\section{Introduction}\label{sec:introduction}
Modern vehicles are complex cyber-physical systems composed of numerous Electronic Control Units (ECUs) that communicate through in-vehicle networks to support powertrain, chassis, body, infotainment, and advanced driver-assistance functions. The Controller Area Network (CAN) remains one of the most widely deployed automotive communication protocols because of its low cost, robustness, and deterministic arbitration behavior suitable for real-time control applications.

Despite its widespread adoption, CAN was designed with reliability and timing as primary objectives rather than cybersecurity. CAN messages are broadcast to all nodes on the bus, and the protocol provides neither source authentication nor cryptographic freshness protection. As a result, an adversary who gains access to the bus can potentially inject spoofed messages or replay previously captured traffic \cite{koscher2010,miller2015,hoppe2011}. Prior research has demonstrated that such attacks can influence safety-critical vehicle functions, making CAN security an important concern for modern automotive systems.

To address these vulnerabilities, the automotive industry has adopted standards-based security mechanisms such as AUTOSAR Secure Onboard Communication (SecOC) and cybersecurity processes defined by ISO/SAE 21434 \cite{autosar2020secoc,iso21434}. In parallel, numerous intrusion detection systems (IDS) have been proposed to identify malicious behavior from CAN traffic patterns. While these approaches improve attack detection, authenticity, and freshness protection, they generally preserve the static communication structure of conventional CAN networks.

This paper investigates a complementary mitigation strategy based on deterministic runtime-adaptive routing diversity. Instead of relying on a fixed logical communication path, participating ECUs periodically switch among predefined routing patterns selected through a shared deterministic function of synchronized and quantized runtime vehicle inputs. The proposed mechanism operates as a logical routing overlay above the CAN protocol without modifying the underlying physical layer or arbitration process. By continuously varying message propagation paths, the approach aims to reduce the likelihood that previously captured traffic remains valid at future times.

The main contributions of this work are summarized as follows:
\begin{itemize}
\item Deterministic runtime-adaptive routing overlay: A routing framework that enables coordinated switching among predefined logical paths while preserving standard CAN operation.
\item Distributed route selection mechanism: A synchronization approach that allows ECUs to independently compute the active routing pattern without centralized control.
\item Replay-resistance evaluation: Analytical and simulation-based assessment of routing entropy, routing-index collisions, and replay attack acceptance rates.
\item Security–robustness trade-off analysis: Investigation of the impact of synchronization tolerance on replay resistance and false message rejection.
\end{itemize}


Simulation results demonstrate that the proposed routing overlay reduces replay acceptance compared with a static-routing baseline while maintaining acceptable synchronization robustness. The mechanism is not intended to replace cryptographic authentication or freshness protection, but rather to serve as an additional defense-in-depth layer for in-vehicle cybersecurity architectures.

\section{Related Work}
\label{sec:related_work}
Automotive cybersecurity research has extensively documented
the security limitations of Controller Area Networks (CAN).
Because CAN does not provide native sender authentication
or message freshness protection, attackers with bus access
can inject or replay messages that may be accepted by
receiving ECUs \cite{koscher2010,miller2015,hoppe2011}.
Previous studies have demonstrated that such attacks can
influence vehicle functionality and safety-relevant behavior.

A significant body of research has focused on intrusion
detection systems (IDS) for CAN networks. Statistical,
machine-learning, and deep-learning approaches have been
proposed to detect anomalous traffic patterns and malicious
message injections \cite{wu2019survey,hanselmann2019canet,
amato2021can,wei2023attention,khan2023ids,bari2023comparative}.
While these methods can effectively identify suspicious
activity, they generally operate after malicious traffic
has already been placed on the network.

Another major research direction involves cryptographic
protection mechanisms. AUTOSAR Secure Onboard Communication
(SecOC) provides message authentication and freshness
verification, while ISO/SAE 21434 establishes cybersecurity
engineering processes for vehicle development
\cite{autosar2020secoc,iso21434}.
These approaches can improve communication
authenticity and integrity but may increase implementation
complexity and resource requirements.

Unlike detection- or authentication-focused approaches,
this work investigates routing diversity as a complementary
mitigation mechanism. The proposed overlay introduces
deterministic runtime-adaptive logical routing paths while
preserving the CAN physical layer and arbitration process.
Rather than replacing cryptographic protections, the
approach is intended to provide an additional
structural-diversity layer within a defense-in-depth
automotive cybersecurity architecture. To the authors' knowledge, comparatively little attention has been given to routing diversity as a replay-mitigation mechanism in CAN-based vehicle networks.

\section{THREAT MODEL AND SYSTEM OVERVIEW}\label{sec:system_overview}

\subsection{Threat Model}\label{threat_model}

The proposed routing overlay is evaluated under a realistic automotive attacker model in which an adversary has access to a CAN bus segment and can observe, inject, and replay network traffic. Similar capabilities have been demonstrated in prior automotive cybersecurity studies \cite{koscher2010,miller2015}. The attacker may record legitimate CAN frames and retransmit them at a later time in an attempt to influence vehicle behavior.
The experimental threat model is limited to delayed replay of previously valid CAN traffic.

Participating ECUs reside on isolated physical subnets connected via zonal gateways. They store predefined routing tables, execute a synchronized routing-index algorithm, and maintain bounded time synchronization $W$. Destination ECUs verify logical routing paths dynamically, rejecting any direct subnet injections that bypass intermediate gateway hops. The CAN physical layer remains unchanged, and attackers are assumed not to possess routing-table contents or firmware-level access to participating ECUs.

This study focuses primarily on replay attacks because replayed messages often preserve valid identifiers, payload structures, and timing characteristics, making them difficult to distinguish without freshness or routing-state validation.
A replay attack occurs when a previously valid CAN frame is retransmitted after a delay in an attempt to satisfy receiver expectations. Under conventional static routing, replayed messages may remain valid if timing and message structure are preserved. The proposed routing overlay aims to reduce this acceptance probability by continuously varying the expected logical message path.

In the evaluated classical CAN layout, routing metadata occupies two payload bytes: one byte for the current-hop source ECU identifier and one byte for the next-hop destination ECU identifier. Thus, six of the eight classical CAN payload bytes remain for application data. The routing index itself is not transmitted; participating ECUs compute the active routing state locally using synchronized inputs and time. 

The proposed mechanism applies to message flows whose participating ECUs are provisioned with the required routing-table entries. Memory usage increases with the number of configured flows and predefined routing patterns; flow-selection policy is a system-integration decision outside the scope of this study.

\subsection{System Overview}\label{system_overview}
The proposed architecture operates as a logical routing overlay above a standard CAN network. No modifications are made to CAN frame arbitration, physical signaling, or bus access mechanisms. Instead, participating ECUs maintain multiple predefined routing patterns for selected communications.
To support routing decisions, each transmitted message includes lightweight source and destination ECU identifiers within the payload. These identifiers enable participating ECUs to determine whether they are part of the currently active logical route.
Routing patterns are selected using a shared deterministic routing-index function:

\begin{equation}
I(t)
=
H(x_1,x_2,\ldots,x_k,t)
\bmod N
\label{eq:routing_index}
\end{equation}

where $x_1$,$x_2$,…,$x_k$ represent synchronized and quantized runtime vehicle inputs, $t$ is the current time interval, and $N$ is the number of predefined routing patterns. Because all ECUs evaluate the same function using the same inputs, the active routing pattern can be selected independently without centralized coordination.

\subsection{Logical Message Forwarding}\label{logical_message_forwarding}

\begin{figure}[t]
\centering
\includegraphics[width=0.8\columnwidth]{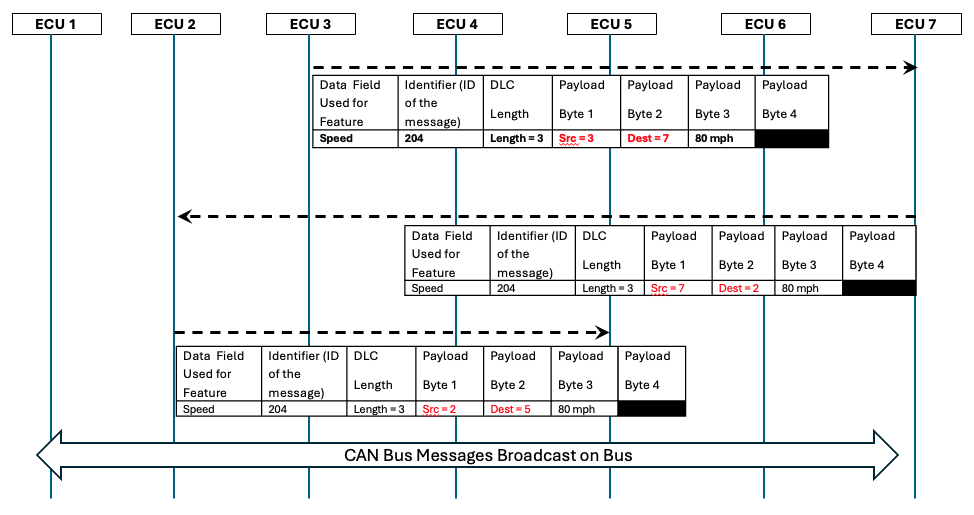}
\caption{Example logical multi-hop message propagation between ECUs under the proposed routing overlay.}
\label{fig:routing_flow}
\end{figure}

When a source ECU transmits a message, participating ECUs evaluate the active routing pattern and forward the frame only when they correspond to the expected next hop. The process continues until the destination ECU receives the message through the active route sequence. Fig.~\ref{fig:routing_flow} illustrates the forwarding process. By varying logical message paths over time, previously captured traffic becomes less likely to satisfy current routing expectations.


\section{DETERMINISTIC RUNTIME-ADAPTIVE ROUTING OVERLAY}\label{sec:routing_overlay}



\subsection{Routing Table Structure}

The proposed overlay introduces a logical routing layer above a conventional CAN network. For selected source-destination communications, participating ECUs store multiple predefined routing patterns that define alternative message propagation paths. Each routing entry consists of a routing index, message identifier, source ECU, destination ECU, and an ordered sequence of intermediate ECUs.
Table~\ref{tab:routing_table} illustrates an example routing table for a representative source-destination communication. Multiple routing patterns are available for a given source-destination communication pair, enabling the active logical path to change over time while preserving deterministic operation. Routing tables are generated offline and distributed to participating ECUs prior to deployment. 
Replay decisions follow the active route sequence, as illustrated in Table I.


\begin{table}[t] \centering \caption{EXAMPLE ROUTING-TABLE ENTRIES AND REPLAY DECISION CASES FOR THE ECU 3 TO ECU 5 FLOW} \label{tab:routing_table} \begin{tabular}{c c p{0.25\linewidth} p{0.28\linewidth}} \hline \textbf{Index} & \textbf{Msg. ID} & \textbf{Route sequence} & \textbf{Example decision} \\ \hline 0 & 204 & $3 \rightarrow 7 \rightarrow 2 \rightarrow 5$ & Legitimate capture at $t_0$: accepted \\ \hline 1 & 204 & $3 \rightarrow 9 \rightarrow 8 \rightarrow 5$ & Replay of index 0 at $t_0+50$: rejected \\ \hline 2 & 204 & $3 \rightarrow 4 \rightarrow 6 \rightarrow 5$ & Legitimate frame: accepted if within $W$ \\ \hline \end{tabular} \end{table}

Conventional CAN communication relies on arbitration identifiers for frame prioritization and does not provide explicit source or destination addressing. To support logical routing decisions, the proposed overlay reserves a small portion of the payload for routing-related metadata.
To support forwarding decisions, source and destination ECU identifiers are embedded within reserved CAN payload bytes, allowing intermediate ECUs to evaluate routing information without modifying arbitration identifiers.

\subsection{Logical Multi-Hop Forwarding}
Although CAN frames remain physically broadcast to all nodes, message propagation is controlled through a logical forwarding process.
When a source ECU transmits a message, all participating ECUs determine the currently active routing pattern. An ECU forwards the received message only if it corresponds to the expected next hop in the active route sequence. Prior to retransmission, the forwarding ECU updates the source field and determines the next hop according to the active routing table entry.

The forwarding process continues until the message reaches the destination ECU. Consequently, message acceptance depends not only on message contents but also on correct progression through the expected logical route.
This mechanism introduces path diversity while remaining generally compatible with standard CAN operation.

\subsection{Synchronization and Implementation Considerations}
Correct route selection requires participating ECUs to maintain bounded synchronization. This can be achieved using periodic synchronization messages, gateway-distributed timing, GPS-based references, or predefined switching schedules.
To accommodate small timing differences, receivers may apply an acceptance window that allows limited synchronization mismatch. The impact of this tolerance on replay resistance and false message rejection is evaluated in Section~\ref{sec:methodology} and Section~\ref{sec:results}.
The proposed overlay is intended to complement existing vehicle cyber-security mechanisms, including AUTOSAR SecOC, intrusion detection systems, secure boot, and hardware security modules. Its purpose is to introduce structural diversity in message propagation rather than replace cryptographic authenticity or freshness protection.

\enlargethispage{-\baselineskip}
\section{SECURITY ANALYSIS}\label{sec:security_analysis}

This section analyzes how runtime-adaptive routing affects replay feasibility and routing predictability. The objective is not to provide cryptographic guarantees, but rather to evaluate how routing diversity influences the probability that previously observed traffic remains valid under future network conditions.

\subsection{Routing Index Entropy}
The unpredictability of routing-pattern selection is characterized using Shannon entropy. The routing-index entropy is defined as \eqref{eq:entropy}, where $N$ is the total number of available routing patterns and $P(I=j)$ is the probability that routing index $j$ is selected.

\begin{equation}
H(I) =
-\sum_{j=0}^{N-1}
P(I=j)\log_{2}P(I=j)
\label{eq:entropy}
\end{equation}

Maximum entropy occurs when all routing patterns are selected with equal probability, yielding $H_{\max} = \log_{2}N$. Higher entropy corresponds to more uniform route utilization and lower route-selection predictability.

\subsection{Replay Acceptance Model}
A replay attack is modeled as an adversary recording a valid message at time $t_0$ and retransmitting it at a later time $t_1$.
For a replayed frame to be accepted, the routing state that was active when the message was recorded must remain valid at the replay instant. Assuming routing patterns are approximately uniformly selected from a set of $N$ predefined routes, the replay acceptance probability can be approximated as 

\begin{equation}
P_{\mathrm{replay}}
\approx
\frac{1}{N}
\label{eq:replay}
\end{equation}

This approximation assumes approximately uniform routing-pattern selection and serves as a first-order estimate of replay feasibility. Actual replay acceptance additionally depends on synchronization constraints and routing-state validation.

This relationship illustrates the primary security benefit of routing diversity. As the number of available routing patterns increases, the probability that a replayed message encounters the same routing state decreases.

\subsection{Routing Index Repetition}
Because routing indices are selected from a finite set, some route reuse is unavoidable over long observation periods. To characterize routing-index repetition, two collision metrics are evaluated.
The window collision rate measures long-term routing-index reuse over an observation interval of length $T$, 

\begin{equation}
C_{\mathrm{window}}
=
1
-
\frac{\text{Number of Unique Indices}}
     {T}
\label{eq:window_collision}
\end{equation}
This metric captures overall route reuse but does not directly reflect practical replay feasibility.
More relevant to replay resistance is the adjacent collision rate, which measures consecutive routing-index repetition:

\begin{equation}
C_{\mathrm{adj}}
=
\frac{
\text{Number of Consecutive Identical Indices}
}
{T-1}
\label{eq:adj_collision}
\end{equation}

\subsection{ Synchronization Sensitivity}
The proposed routing overlay assumes bounded synchronization among participating ECUs. Let $\delta_t$ denote the synchronization offset between communicating nodes. To tolerate small timing discrepancies, receivers may apply an acceptance window $W$.

\begin{equation}
t_{\mathrm{expected}}
\in
[t_{\mathrm{observed}}-W,
t_{\mathrm{observed}}+W]
\label{eq:acceptance_window}
\end{equation}

Increasing $W$ improves robustness to synchronization jitter and reduces false message rejection. However, a larger acceptance window may also increase replay acceptance by allowing previously valid routing states to remain acceptable for longer periods. Consequently, synchronization tolerance introduces a trade-off between communication reliability and replay resistance.

\section{EXPERIMENTAL METHODOLOGY}\label{sec:methodology}
\subsection{Simulation Configuration}
The overlay was evaluated using a custom synchronous discrete-time state simulator. One simulation step represents one route-selection interval rather than CAN bit time; arbitration latency, error handling, and detailed bus utilization are not modeled. The representative network contains 10 ECUs, with ECU 3 as the source and ECU 5 as the destination for the illustrated flow. The experiment uses 500 routing-table states over 2,500 steps, with matching offline-generated tables stored by participating ECUs.
Table~\ref{tab:sim_params} summarizes the principal simulation parameters used throughout the evaluation.

Replay is modeled by recording a valid frame at step $t_0$ and retransmitting it after a fixed delay of $d=50$ steps. The replay is accepted only if the recorded route information is consistent with the active routing-table state at replay time and falls within the configured synchronization window $W$.

\begin{table}[t]
\caption{Simulation Parameters}
\label{tab:sim_params}
\centering
\begin{tabular}{lc}
\hline
Parameter & Value \\
\hline
Number of ECUs & 10 \\
Routing Patterns ($N$) & 500 \\
Simulation Duration & 2500 Steps \\
Routing Hops & 3 \\
Replay Delay & 50 Steps \\
Acceptance Window ($W$) & 0--5 \\
Timing Jitter ($\sigma$) & 0--1 \\
\hline
\end{tabular}
\end{table}

For each simulation step, all participating ECUs compute the active routing index using the same deterministic routing-index function and quantized input vector. The representative protected flow uses message ID 204 from ECU 3 to ECU 5, consistent with Table~\ref{tab:routing_table}. Each routing pattern defines an ordered logical path, such as 3→7→2→5 or 3→9→8→5.

The input vector $x_1,\ldots,x_4$ consists of cumulative wheel ticks, transmission gear state, engine speed quantized in 50 RPM blocks, and global gPTP epoch timestamps. These synthetic, quantized inputs evaluate routing behavior rather than a specific vehicle platform.


\subsection{Replay Attack Modeling}
Replay attacks were modeled by allowing an adversary to observe network traffic, capture valid CAN messages, and retransmit them after a fixed delay. A replayed message was considered successful only if the routing state at replay time satisfied the active routing pattern and synchronization requirements.


\subsection{False Rejection Evaluation}
Synchronization imperfections were modeled using timing jitter with standard deviation $\sigma$. A false rejection occurs when a legitimate message is rejected because participating ECUs are not sufficiently synchronized.



\section{RESULTS}\label{sec:results}

\subsection{Routing Index Entropy}
The routing-index distribution was evaluated over the 2500-step simulation period using the Shannon entropy metric defined in Section~\ref{sec:security_analysis}. For $N=500$ routing patterns, the measured entropy was approximately 8.8 bits, compared with the theoretical maximum of $H_{max}=log_2(500) \approx 8.97$ bits. 

The measured entropy was within 2\% of the theoretical maximum, indicating near-uniform utilization of the available routing patterns and low route-selection predictability.


\subsection{Replay Attack Acceptance}
Replay acceptance was evaluated for different synchronization tolerance windows and timing-jitter conditions. A static-routing baseline was also considered for comparison.
Under static routing, replayed frames remain valid unless additional protection mechanisms such as freshness verification are present. Consequently, replay acceptance is assumed to approach unity for the baseline configuration. This baseline is a simulation construct used solely for comparative purposes; it does not describe the security architecture of any production vehicle or account for independent protections such as SecOC or IDS that may be present in fielded systems. The phrase 'approach unity' reflects the absence of additional modeled protections within the simulation and is not intended as empirical characterization of any system in deployment.

\begin{figure}[t]
\centering
\includegraphics[width=0.8\columnwidth]{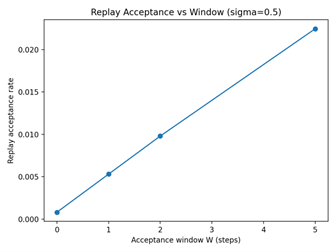}
\caption{Replay attack acceptance probability under varying synchronization acceptance windows.}
\label{fig:replay_acceptance}
\end{figure}

For the proposed runtime-adaptive routing overlay, replay acceptance decreased substantially. Replay acceptance results are summarized in Table~\ref{tab:replay_results}. With zero synchronization tolerance ($W=0$), the replay acceptance rate was approximately 0.002, closely matching the analytical expectation of $1/N$ for $N=500$ routing patterns. As shown in Table~\ref{tab:replay_results}, replay acceptance increases as the synchronization window becomes larger since previously valid routing states remained acceptable for longer periods.
Even under the most permissive synchronization conditions evaluated, replay acceptance remained below 3\%, representing a substantial improvement compared with the static-routing baseline.
Fig.~\ref{fig:replay_acceptance} illustrates the relationship between synchronization tolerance and replay acceptance probability.

\begin{table}[!t]
\caption{Replay Acceptance Rate}
\label{tab:replay_results}
\centering
\begin{tabular}{ccc}
\hline
$W$ & $\sigma$ & Replay Rate \\
\hline
0 & 0.0 & 0.002 \\
1 & 0.5 & 0.005 \\
2 & 0.5 & 0.010 \\
5 & 1.0 & 0.025 \\
\hline
\end{tabular}
\end{table}

\subsection{False Rejection Rate}
False rejection was evaluated to quantify the impact of synchronization mismatch on legitimate traffic. False rejection results are summarized in Table~\ref{tab:false_rejection}. As expected, strict synchronization requirements reduced replay acceptance but increased rejection of valid messages when timing jitter was present. Introducing a modest acceptance window improved communication robustness.

\begin{table}[!t]
\caption{False Rejection Rate}
\label{tab:false_rejection}
\centering
\begin{tabular}{ccc}
\hline
$W$ & $\sigma$ & False Rejection \\
\hline
0 & 0.5 & 8\% \\
1 & 0.5 & 2\% \\
2 & 0.5 & $<1\%$ \\
5 & 1.0 & $<0.5\%$ \\
\hline
\end{tabular}
\end{table}

\begin{figure}[!t]
\centering
\includegraphics[width=0.8\columnwidth]{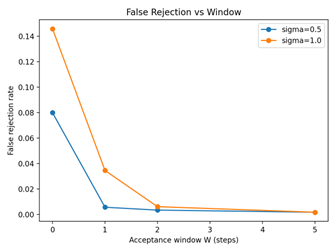}
\caption{False rejection rate under varying synchronization tolerance and timing jitter.}
\label{fig:false_rejection}
\end{figure}

These results indicate that relatively small synchronization windows can substantially reduce false rejection while maintaining low replay acceptance. 
Fig.~\ref{fig:false_rejection} shows the effect of synchronization tolerance on false rejection behavior.

These observations highlight the trade-off between replay resistance and synchronization robustness, with $W=1$ and $W=2$ providing the most favorable operating region.

\subsection{Routing Index Repetition Behavior}
Collision behavior was evaluated using both the window-collision and adjacent-collision metrics defined in Section~\ref{sec:security_analysis}.
Because the observation window ($T=2500$) exceeded the number of routing patterns ($N=500$), long-term routing-index reuse was expected and resulted in a window-collision rate of approximately 80\%. However, this behavior has limited impact on practical replay feasibility because replay attacks typically depend on short-term routing-state repetition.

The observed adjacent-collision rate was approximately 0.08\%, remaining below the theoretical expectation of $1/N=0.2\%$. This indicates that successive routing states rarely repeat, reducing the likelihood that a replayed frame encounters the same routing conditions as the original transmission.

Collectively, these results indicate that runtime-adaptive routing substantially decreases replay feasibility while maintaining acceptable synchronization robustness under the evaluated operating conditions.



\section{DISCUSSION}\label{sec:discussion}
\subsection{Security Implications}
The simulation results indicate that introducing runtime-adaptive routing diversity can reduce replay feasibility under the evaluated simulation conditions. Under static routing, an attacker primarily needs to reproduce the expected message identifier, payload structure, and timing characteristics. In contrast, the proposed overlay introduces an additional dependency on the active routing state, requiring a replayed message to align with the currently selected routing pattern.

Consequently, a previously valid message may no longer satisfy receiver expectations when retransmitted at a later time. This additional routing-state dependency increases the difficulty of successful replay attacks and provides a complementary layer of protection beyond traditional CAN communication mechanisms.

It is important to note that the proposed routing overlay is not intended to provide cryptographic security. It is best viewed as a complementary mechanism to freshness and authentication frameworks such as AUTOSAR SecOC \cite{autosar2020secoc}. Instead, it aims to reduce attack feasibility through controlled route diversity and time-varying communication behavior.

\subsection{Security-Reliability Trade-Off}
The results demonstrate a trade-off between replay resistance and synchronization robustness. Smaller receiver acceptance windows provide stronger replay resistance by enforcing stricter agreement on the active routing state. However, these same constraints increase the likelihood that legitimate messages will be rejected when timing jitter is present.

Conversely, increasing synchronization tolerance reduces false rejection and improves communication robustness, but also increases replay acceptance because routing-state validation becomes less restrictive. The evaluated results suggest that moderate synchronization windows provide a practical balance, maintaining low replay acceptance while avoiding excessive rejection of legitimate traffic.

This trade-off is particularly relevant for automotive systems where both cybersecurity and dependable real-time communication must be considered simultaneously.

\subsection{Implementation Considerations}
Although the proposed overlay preserves the CAN physical layer and arbitration mechanism, practical deployment introduces several implementation challenges.

First, routing tables must be stored within participating ECUs, creating additional memory requirements that scale with the number of supported routing patterns. Second, logical multi-hop forwarding increases bus traffic because intermediate ECUs retransmit messages along the selected routing path. For a route containing $k$ forwarding stages, each application-level message generates approximately $k$ additional CAN transmissions.

Furthermore, forwarding operations introduce additional end-to-end latency. Any future application of this research would be scoped through standard engineering and safety-validation processes; scope determination is outside the scope of this paper.

Consequently, routing diversity is most appropriately treated as a configurable overlay applied to message flows for which participating ECUs are provisioned with the required routing-table entries, rather than as an assumed default for all CAN traffic.

\subsection{Limitations}
Several limitations should be acknowledged.

\begin{itemize}
\item The evaluation was performed in simulation and does not include hardware-in-the-loop or vehicle-level testing.
\item The study does not evaluate cryptographic key compromise, persistent denial-of-service attacks, or hardware side-channel attacks.
\item Additional CAN bus utilization and latency effects require further investigation under realistic vehicle traffic conditions.
\item Integration with AUTOSAR SecOC and other production cybersecurity frameworks was not experimentally evaluated.
\item ECU firmware compromise, routing-table or synchronization compromise, live spoofing, denial-of-service, key compromise, and physical-layer attacks are outside scope and require complementary defenses.
\end{itemize}

These limitations suggest that the routing overlay should be viewed as one component within a broader defense-in-depth strategy rather than as a standalone security solution.

\section{CONCLUSION}\label{sec:conclusion}
This paper proposed a deterministic runtime-adaptive routing overlay for Controller Area Network (CAN) systems aimed at reducing replay-attack feasibility without modifying the CAN physical layer or arbitration mechanism. The proposed approach introduces logical route diversity by allowing participating Electronic Control Units (ECUs) to dynamically select among predefined routing patterns using synchronized runtime information. As a result, message acceptance becomes dependent not only on payload content and timing but also on the active routing state.

Simulation results demonstrated that the proposed overlay substantially reduces replay acceptance when compared with a static-routing baseline. With 500 predefined routing patterns, routing-index entropy approached its theoretical maximum, indicating near-uniform route utilization. Replay acceptance decreased to approximately 0.2\% under strict synchronization conditions and remained below 3\% under the most permissive synchronization window evaluated. The results also showed that modest synchronization tolerance can reduce false rejection while preserving most of the replay-resistance benefits.

The proposed mechanism is not intended to replace cryptographic authentication or freshness protection mechanisms such as AUTOSAR SecOC. Instead, it provides an additional structural-diversity layer that may strengthen defense-in-depth automotive cybersecurity architectures. 
Future work will evaluate hardware-in-the-loop validation, vehicle-network effects, bus load, latency, and standards-based authentication integration.

\bibliographystyle{IEEEtran}
\bibliography{references}

\textbf{Disclaimer}: 
1. This publication was written based on the state-of-the-art automated driving technology at the time of publication. As such, it is not a complete work and its contents should be continuously revisited and revised whenever advances are made in the areas of social acceptance, technology and legislation. For the avoidance of doubt, this publication is not intended and shall not be construed to establish, create, or deem to have created a maximum requirement regarding dependability for automated driving, nor shall it prevent, hinder, or restrict any author's employer, or any company affiliated with an author's employer, from deviating in any way from the content of this publication. In case any statement of this publication may be perceived as conflicting with this vision, the non-binding vision shall always prevail.

2. This paper presents a research concept evaluated solely through offline simulation. Nothing in this publication describes, represents, or warrants the cybersecurity architecture, configuration, or performance of any Torc Robotics production vehicle or system currently in commercial service. Vehicle input parameters, routing-table structures, message identifiers, and other implementation details presented herein are illustrative and used solely to support the described simulation; they should not be construed as a description of any specific deployed system. The mechanism described is intended as a complement to, and not a substitute for, cryptographic authentication and other defense-in-depth measures, and this publication should not be construed as an admission regarding the adequacy or inadequacy of the cybersecurity posture of any particular vehicle or system.

\end{document}